\documentclass{sigchi}

\usepackage{balance}       
\usepackage{graphics}      
\usepackage[T1]{fontenc}   
\usepackage{txfonts}
\usepackage{mathptmx}
\usepackage[pdflang={en-US},pdftex]{hyperref}
\usepackage{color}
\usepackage{booktabs}
\usepackage{textcomp}
\usepackage{eurosym}
\usepackage{enumitem}

\usepackage{microtype}        
\usepackage{ccicons}          

\usepackage{todonotes}

\def\plaintitle{Evaluation of a Search Interface for Preference-Based Ranking - Measuring User Satisfaction and System Performance}

\def\emptyauthor{}
\def\plainkeywords{Search interface, Information filtering, Preference-based ranking, Weighted facets, Evaluation of whole sessions}

\makeatletter
\def\url@leostyle{%
  \@ifundefined{selectfont}{
    \def\UrlFont{\sf}
  }{
    \def\UrlFont{\small\bf\ttfamily}
  }}
\makeatother
\def\pprw{8.5in}
\def\pprh{11in}

\definecolor{linkColor}{RGB}{6,125,233}
\hypersetup{%
  pdftitle={\plaintitle},
  pdfauthor={\emptyauthor},
  pdfkeywords={\plainkeywords},
  pdfdisplaydoctitle=true, 
  bookmarksnumbered,
  pdfstartview={FitH},
  colorlinks,
  citecolor=black,
  filecolor=black,
  linkcolor=black,
  urlcolor=linkColor,
  breaklinks=true,
  hypertexnames=false
}

\begin{document}
	
	\CopyrightYear{2018}
	\setcopyright{acmlicensed}
	\conferenceinfo{NordiCHI'18,}{September 29-October 3, 2018, Oslo, Norway}
	\isbn{978-1-4503-6437-9/18/09}
	\acmPrice{\$15.00}
	\doi{https://doi.org/10.1145/3240167.3240170}

\title{\plaintitle}

\numberofauthors{3}
\author{%
  \alignauthor{Dagmar Kern\\
    \affaddr{GESIS - Leibniz-Institute for the Social Science}\\
    \affaddr{Cologne, Germany}\\
    \email{dagmar.kern@gesis.org}}\\
  \alignauthor{Wilko van Hoek\\
    \affaddr{Bonn, Germany}\\
    \email{wilko.vanhoek@gmail.com}}\\
  \alignauthor{Daniel Hienert\\
    \affaddr{GESIS - Leibniz-Institute for the Social Science}\\
    \affaddr{Cologne, Germany}\\
    \email{daniel.hienert@gesis.org}}\\
}

\makeatletter
\def\@copyrightspace{\relax}
\makeatother

\maketitle

\begin{abstract}
  	Finding a product online can be a challenging task for users. Faceted search interfaces, often in combination with recommenders, can support users in finding a product that fits their preferences. However, those preferences are not always equally weighted: some might be more important to a user than others (e.g. red is the favorite color, but blue is also fine) and sometimes preferences are even contradictory (e.g. the lowest price vs. the highest performance). Often, there is even no product that meets all preferences. In those cases, faceted search interfaces reach their limits. In our research, we investigate the potential of a search interface, which allows a preference-based ranking based on weighted search and facet terms. We performed a user study with 24 participants and measured user satisfaction and system performance. The results show that with the preference-based search interface the users were given more alternatives that best meet their preferences and that they are more satisfied with the selected product than with a search interface using standard facets. Furthermore, in this work we study the relationship between user satisfaction and search precision within the whole search session and found first indications that there might be a relation between them.
\end{abstract}


\category{H.5.2.}{Information Interfaces and Presentation
	(e.g. HCI)}{User Interfaces -- Evaluation/methodology, Graphical user interfaces (GUI)} 
\category{H.3.3.}{Information Storage and Retrieval }{Information Search and Retrieval -- Query formulation}

\section{Introduction}
The number of consumers who browse or buy products online grows further \cite{internetretailer:venue}, and they are all facing the challenge to choose the best option out of a huge set of alternatives. Web shops and service providers (e.g. hotel booking services or apartment finders) try to support users in finding the right product. One well-established method for that is to provide search facets in the user interface. \cite{Hearst:2002:FFW:567498.567525}. Facets allow users to filter products by predefined categories or features. In this way, users can exclude products they are not interested in and obtain a smaller and more manageable number of alternatives. An overview about faceted search in general is given by \cite{tunkelang:doi:10.2200/S00190ED1V01Y200904ICR005, Wei:2013:SFS:2481562.2481564}. Studies have shown that search interfaces providing facets are considered as intuitive and easy to use, see e.g. \cite{Hearst:2002:FFW:567498.567525}. Furthermore, they provide a high level of control and transparency \cite{Yee:2003:FMI:642611.642681}. However, user preferences can not always be encoded into a boolean logic. Sometimes, some features are not mandatory but nice to have and some of them are considered more important than others. But, using facets means that all selected facet terms are mandatory and equally weighted. Selecting less important facet terms may remove potentially interesting products unintentionally, while specifying only a few criteria may lead to a too large result set where interesting alternatives are not obvious \cite{Voigt:2012:WFB:2305484.2305509}. 

To counterbalance disadvantages of facet search, recommender systems \cite{Ricci2011} are often additionally applied to suggest possible alternatives that might fit users' preferences. In current systems, the recommendations rely on user profiles and the use of automated recommendation techniques \cite{Ricci2011}. In general, common commercial recommender systems offer no or little options for users to explicitly influence the recommendations, leaving them no opportunity to express their preferences. However, in the research literature, one can find some evidence that, regarding the user satisfaction, it is beneficial to put users in control of their recommendations (e.g. \cite{bostandjiev2012tasteweights, Harper:2015:PUC:2792838.2800179}) and of the ranking process of search results in general (e.g. \cite{Frei93effectivenessof}). One possibility to give users control over recommendation and ranking is to let them weight terms according to their preferences (e.g. \cite{loepp2015blended, Voigt:2012:WFB:2305484.2305509}). From a user perspective, these systems have been evaluated as very helpful, and they are able to increase user satisfaction (e.g. \cite{Chen2012, Harper:2015:PUC:2792838.2800179}). However, little attention has been given so far considering both sides to evaluate a search system -- user feedback as well as system performance and their relationship to one another. We want to close this gap by providing new insights about a search interface that uses preference-based ranking in the form of weighted facet terms. We want to know if the user's perceived satisfaction and system support can be backed up by analyzing system performance measures. For that purpose, we tracked the changes in recall and precision over the course of whole search sessions. 

\section{Related Work}
Most of the existing commercial search interfaces using facets still offer few opportunities for the user to adapt the search query explicitly to her preferences. However, in research, several attempts have been made to include user preferences in \textit{product search}. For example, Stolze \cite{stolze2000soft} proposed a soft navigation approach for finding products in an electronic product catalog. He distinguished between hard and soft constraints and allows weighting the importance of product features. The proposed system requires from the user to learn a rule based syntax. Therefore, it is stated to be more suitable for frequent users. However, in the research field of \textit{search interfaces} using facets, there are surprisingly few approaches focusing on allowing the user to express their preferences through weighting facet terms. Han et al. \cite{han2016user} focus in their people search system on using slider-based facets to specify the importance of three predefined categories. Their results show that the users consistently interact with the sliders to fine-tune the result ranking to achieve a better ranking. An approach very similar to ours is presented by Voigt et al. \cite{Voigt:2012:WFB:2305484.2305509}. They distinguish in their VisBoard application between must-have and optional facet terms that can be weighted by the user with drag-and-drop in a configuration area. They utilized this approach for the specific task of selecting a visualization component based on its characteristics. Unfortunately, they only performed a preliminary user study with five participants showing the general potential of this approach to support the user expressing her preferences. 

In the field of \textit{recommender systems}, a lot of research approaches show the benefit of including user's preference to increase recommendation accuracy, user satisfaction and user experience. Critiquing-based recommender systems (e.g. see \cite{Chen2012, Bruke}) allow users to interactively criticize recommendations and thus put the users in control of finding a product that fits their preferences. With the possibility to weight critiques (e.g. "less expensive" or "compromise distance"), a user can perform trade-offs while searching for products, a concept which can still be very rarely found in commercial systems. Studies have shown that the critiquing-based approach leads to a higher decision accuracy compared to non critiquing-based systems such as a ranked list with one ranking criteria at a time \cite{Pu:2005:ITS:1064009.1064038, Pu:2004:EES:988772.988804}. In TasteWeight \cite{ bostandjiev2012tasteweights} users can adjust their taste interactively during a recommendation session via slider-weights components. Thus, they weight suggested recommendation terms and influence directly the recommendations. Results of a user study showed a positive effect on user satisfaction. Harper et al. {\cite{Harper:2015:PUC:2792838.2800179} also put the user in control of her recommendations by providing means for tuning system generated recommendations according to her preferences (e.g., "show more popular items"). Results of their user study show that if users have control over a recommender system, they evaluate suggested recommendations more positively than automatically generated recommendations. A study with SetFusion \cite{Parra:2014:SYW:2557500.2557542}, an interactive system that allows weighting the influence of three different recommender algorithms, also shows that due to the interaction and visualization, the users have a  greater sense of perceived control and transparency. The approaches above focus primarily on ranking or criticizing a generated set of recommendations and recommended terms. Loepp et al. \cite{loepp2015blended} in contrast, integrate different recommender algorithms with several interactive filter techniques in one hybrid recommender system. This system allows the selection of hard and soft filter criteria from different facets by the user. The selected facet values can be weighted by the user and serve as input for collaborative and content-based recommender techniques. Results of a user study showed that users feel more in control with the hybrid recommender system than with a standard faceted filtering system. They find the interaction to be more appropriate for generating recommendations.	
	
	In the field of \textit{information retrieval} different concepts have been proposed to involve the user's preferences more interactively in the ranking of search results. A core concept is relevance feedback \cite{rocchio_relevance_1971}, in which the user implicitly or explicitly influences the ranking by marking some result items more relevant than others. Another concept that is strongly related to relevance feedback, is query expansion \cite{carpineto_survey_2012}. Here, user search terms are expanded with additional terms originating from knowledge sources, search results, the document corpus or the user's history. These expanded terms are often assigned a different weight in the overall query, to balance the effect of query expansion. On the user interface side, Frei and Qui \cite{Frei93effectivenessof} allow the user to weight query terms in the context of document retrieval. They showed that weighted queries perform significantly better than Boolean retrieval regarding usefulness and precision. In the context of a digital library, an approach for conditional weighting like preference A is more important than B was formally introduced by \cite{Spyratos2006}. 
	
	In this paper, we build up on different concepts of the works above: allowing users to enter query terms, a recommender for facets, the possibility to weight free and facet terms, a certain fuzziness for specific facets and a highly interactive search interface. We especially concentrate on a thorough evaluation measuring user feedback and system performance based on the whole search session.

	\section{HOTEL SEARCH -- AN EXAMPLE APPLICATION FOR A PREFERENCE-BASED RANKING SYSTEM}\label{paragraph:dataset}
	Searching for a hotel includes a wide variety of facets (hotel features) on the system side as well as some different preferences on the user side. We chose this as a use case for studying the potential of a preference-based ranking system. Our search interface allows optional search terms to enhance the result list with alternatives which do not necessarily fulfill all of the user's preferences. The user can specify which of her preferences have to be matched exactly ("must-have terms") and which preferences are nice to have ("optional terms"). Furthermore, the user can explicitly exclude hotels containing features which are not wanted ("must-not have"). "Must-have" and "must-not have" terms cause a trimming of the result list, while the weighting of "optional terms" influences the ranking so that the best-matched hotels are at the top. 
	
	\begin{figure}[]
	\centering
	\includegraphics[width=1.0\linewidth]{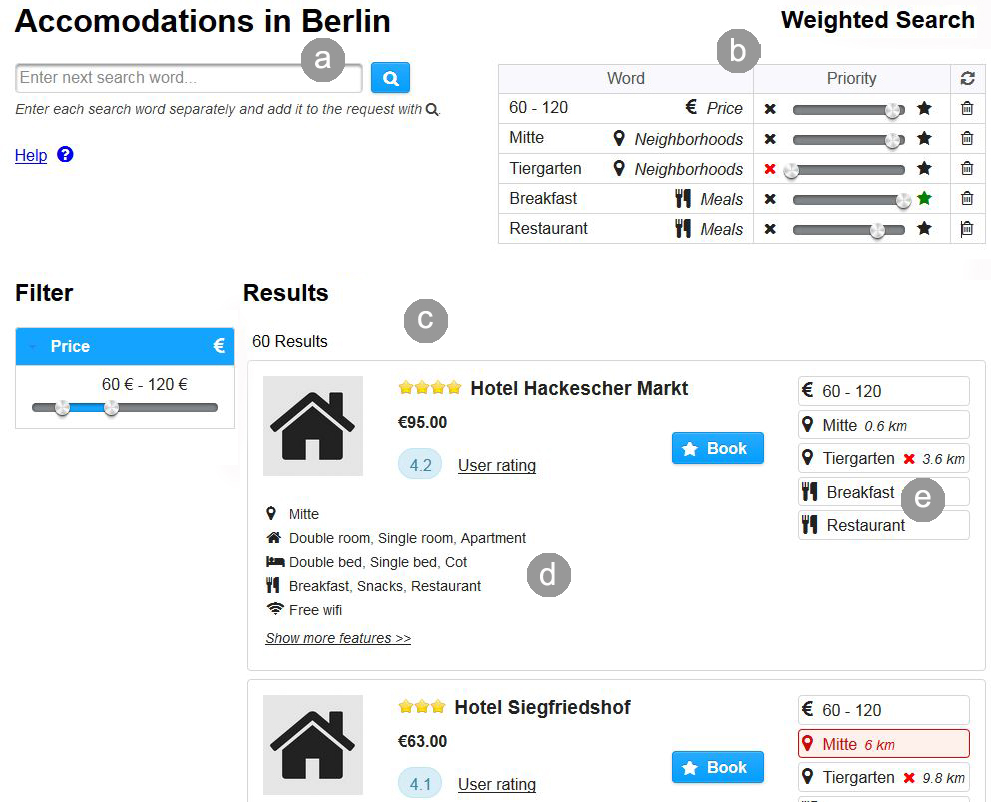}
	
	\caption{The preference-based hotel search interface consists of a) input field, b) weighting area, c) result list, d) list of product features, e) visual feedback on matched or mismatched search terms.}
	\label{fig:mockup}
	\end{figure}
	
	\subsection{User Interface}
	The search interface consists of three main components: input field, weighting area, and search result list. Figure \ref{fig:mockup} shows the implemented interface. In the input field, the user can enter her preferences one after another (Figure  \ref{fig:mockup}a). While typing, a recommender suggests a list of facet terms or facet categories that match the current character string. If the search term does not match a facet term, it is marked as a free text term. In the weighting area (Figure \ref{fig:mockup}b) all search terms which formulate the search query are listed with the possibility to weight each of them. Initially, all search terms have the same almost highest weight. This means that all products are still in the result list. With the provided slider the user can weight the impact of each term. By setting the slider to the minimum of the scale, the search term is considered as a "must-not have" and by setting the slider to the maximum of the scale the search term is considered as a "must-have". With criteria marked as "must-haves" and "must-not haves" the number of results can be decreased. With a click on the recycle bin icon, the search term can be easily removed from the search query. Any interaction in the weighting area leads directly to a new calculation of the search results (Figure \ref{fig:mockup}c) and provides immediate feedback to the user. On the left side of each result item, information on all important hotel features are shown (Figure \ref{fig:mockup}d). On the right side matched and mismatched search terms are visualized (Figure \ref{fig:mockup}e). This gives the user the opportunity to judge the agreement of search results with her preferences and provides transparency over the ranking process. 
	
	\subsection{Determining the Result Set}
	The weights for each search criterion are mapped to floating point numbers between zero and one. This weighting factor is used to determine which hotel will be included in the result set. If the weight is set to one, a hotel will be retrieved if it satisfies the required criterion. In contrary, if the weight is set to 0, the weighting acts as a NOT operator, in which hotels that satisfy the criterion will be eliminated from the result list. This allows users to exclude unwanted criteria. In both cases, hotels will not be included if they lack the given criterion. These cases can be considered as filtering the result set. 
	
	\begin{table}[]
	\centering
	\includegraphics[trim=50 340 250 50,clip,width=1\linewidth]{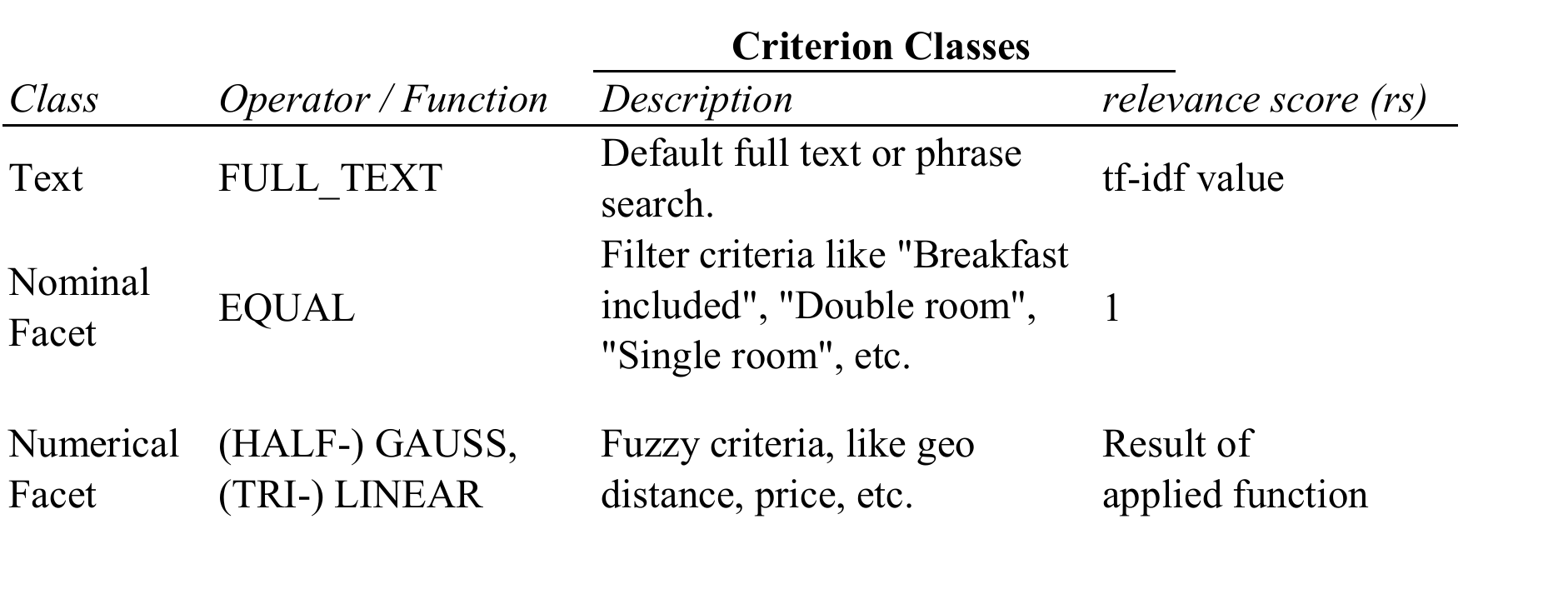}
	\caption{Criterion classes and relevance scores.}
	\label{fig:tab1}
	\end{table}
	
	\subsection{Ranking Model}
	Each hotel in the result lists gets a summed relevance score ($srs$) according to that $srs$ the result list is ordered. The hotel with the highest $srs$ is on top of the list followed by the others in descending order. The summed relevance score for each hotel is computed as $srs$ = $\sum_{i=1}^n w_{Ci} * rs_{Ci}$, whereas (C1-n) are the hotel's criteria the search terms are mapped to. $w$ is the user defined weight for each criterion and $rs$ a relevance score that depends on the criterion class the search term is mapped to (see Table \ref{fig:tab1}). Search terms that cannot be mapped to the given facets are assigned to the class \textit{Text}, for which a tf-idf scoring is applied to calculate a relevance score taking all textual description of the hotel into consideration. A hotel criterion that fulfills a nominal facet (like "breakfast" or "single room") gets a relevance score of 1. For criteria that fulfill numerical facet terms, we suggest using a fuzzy relevance scoring approach whereby hotels that fully match the stated criterion get the highest $rs$ and hotels whose feature fall in a range around the stated value get a lower $rs$ according to the applied function. We propose the usage of a "Gaussian", "linear" or a "tri-linear" scoring function. 
	\begin{figure}[]
	\centering
	\includegraphics[width=1\linewidth]{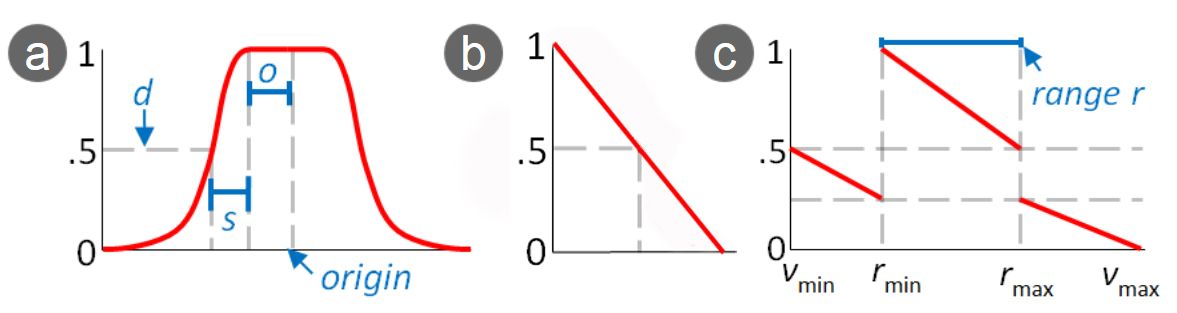}
	\caption{a) Customizable Gaussian function for relevance scoring, b) linear function applied on directed Likert-scales, c) Tri-linear descendant directed scoring function.}
	\label{fig:functions}
	\end{figure}
	Figure \ref{fig:functions}a shows a customizable Gaussian function that can be applied to a criterion for which a single value is given. An optional offset parameter can widen the range where the score will be the maximum. A linear function can be applied where a maximum or minimum is stated (shown in Figure \ref{fig:functions}b), for example, for user ratings . In case a scoring is applied to a criterion given by a range \textit{r} (e.g. "price from \$100-\$120") a tri-linear function can be used which primarily favors the selected range and secondly the border ranges in order of direction (Figure \ref{fig:functions}c). Here, each range is valued differently, so that one range border gets a higher value than the other. In our price example, this means that the lowest price of the selected range gets the highest relevance score ($rs$). The $rs$ decreases linearly until the highest price of the selected range is reached. Prices that do not fell into the price range get lower $rs$ whereby lower prices get higher $rs$ than higher prices, in each case the $rs$ is linearly descending from the lowest to the highest price in each range border.  
	For the first hotel in Figure \ref{fig:mockup}, Table \ref{Tab:example} shows the mapping of the search terms to the hotel criteria, the weights provided by the user, the relevance functions that are used and the calculated relevance score for each criterion. The hotel gets, therefore, a summed relevance score of 3.26. 
	
	\begin{table}[]
	\centering
	\includegraphics[trim=50 560 100 50,clip,width=1\linewidth]{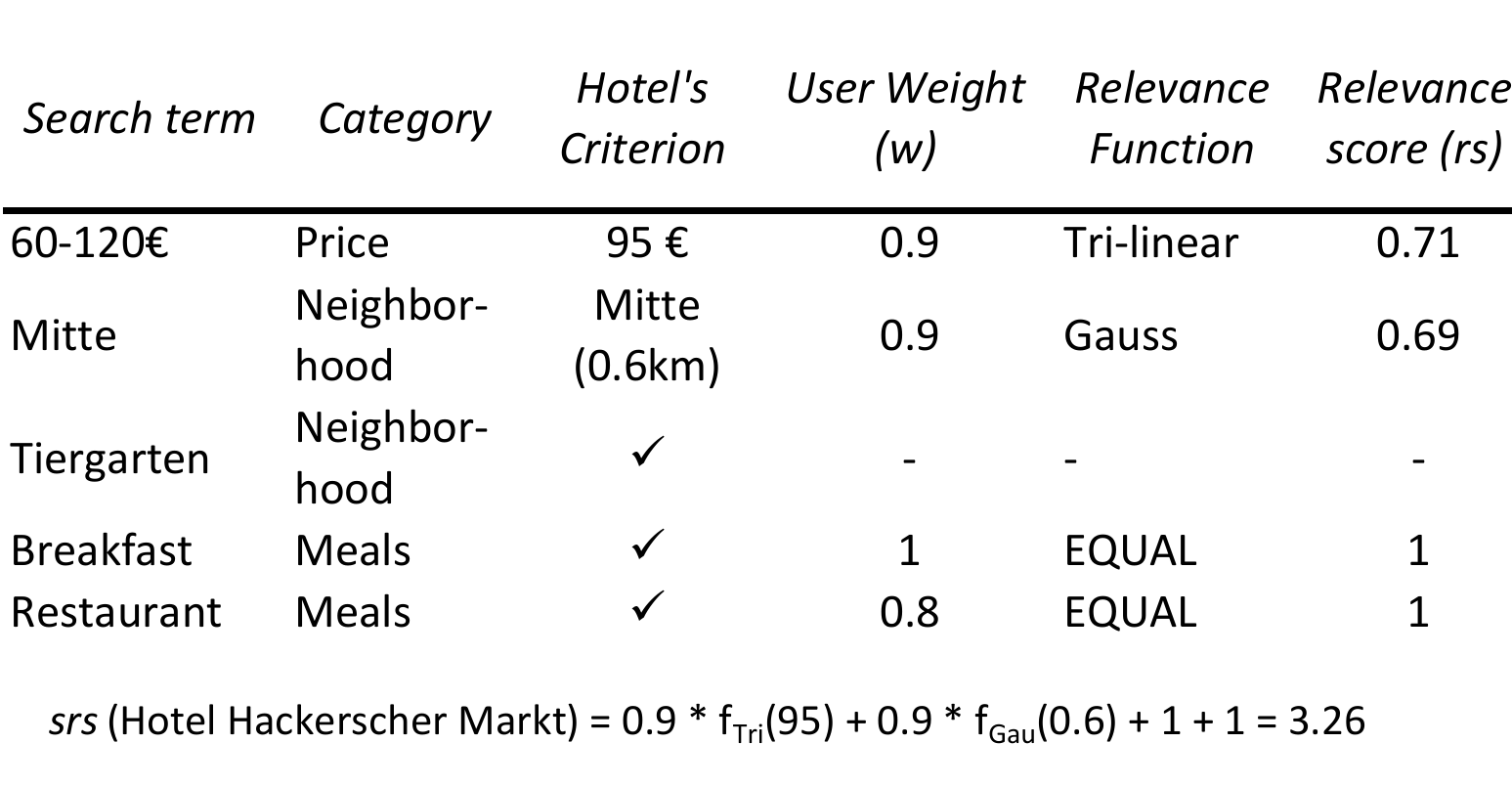}
	\caption{Example mapping from the search terms to the hotel criteria of the first hotel in the results of Figure \ref{fig:mockup} including provided user weights, applied functions and calculated relevance scores.}
	\label{Tab:example}
	\end{table}
	
	\subsection{Implementation}
	For implementing our search interface we utilized Elastic Search\footnote{https://www.elastic.co/de/products/elasticsearch} as a search engine. The software architecture was designed to support various types of data and data sources. The prototype is implemented in Java-EE and provides a generic library and the actual hotel search application. The user interface is a generic JSF web fragment which provides a custom component and java controller (bean)\footnote{The prototype's source code is available here: https://git.gesis.org/iir/preferenced-based-search.}.
	
	The hotel search application contains specific facets with different classes and functions. For the facet "neighborhood", a Gauss scoring (Figure \ref{fig:functions}a) is applied to consider the distance of a hotel to a specified neighborhood. That means hotels that are nearer to the neighborhood indicated in the search query are weighted higher than those farther away. For the "customer rating values" and "hotel stars" linear functions (Figure \ref{fig:functions}b) are used. For the price range, a tri-linear function (shown in Figure \ref{fig:functions}c) is used with a range extended by 20\% to each side of the specified price maximum and minimum. For all nominal facets, an equal value comparison is applied. 
	
	\begin{figure}[]
	\centering
	\includegraphics[width=1\linewidth]{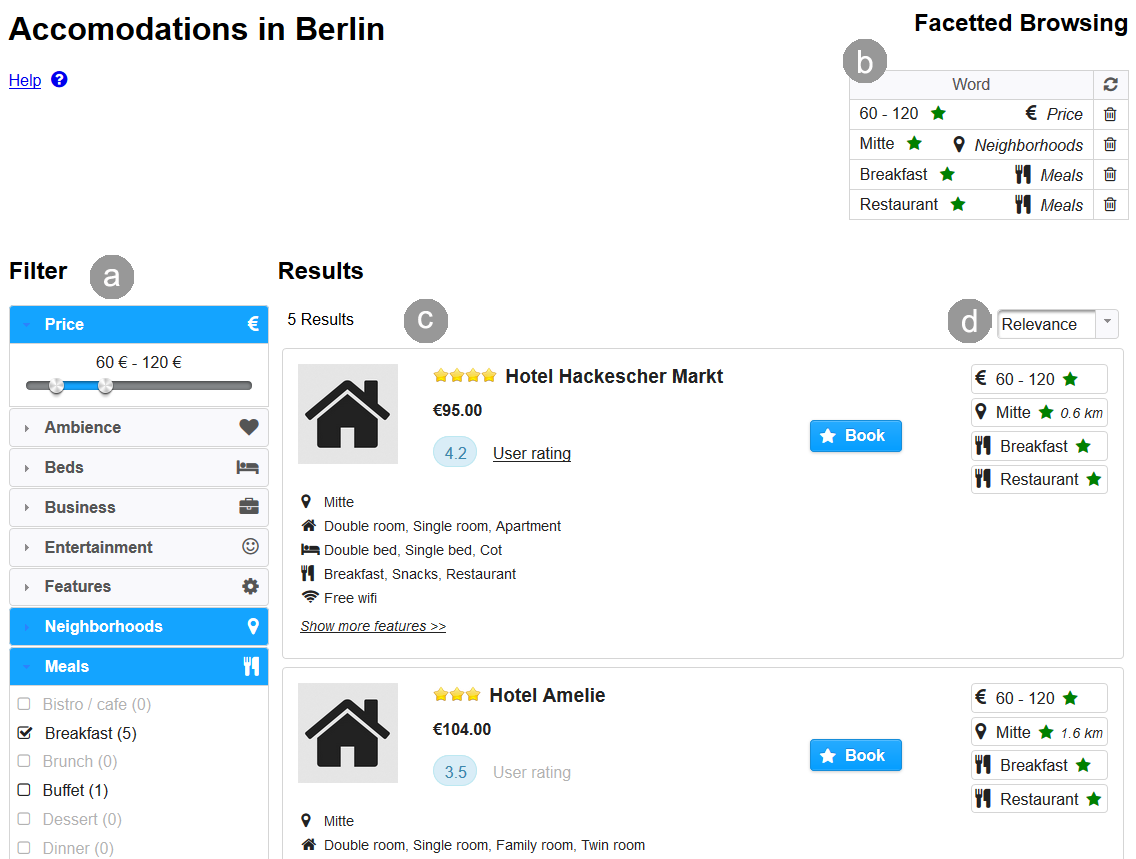}
	\caption{Comparative prototype used in the user study as a representative of a faceted search interface.}
	\label{fig:facet_prototype}
	\end{figure}
	
	\section{Evaluation}
	We performed a lab study to evaluate the presented hotel search interface against a standard search interface using facets (see Figure \ref{fig:facet_prototype}). Additionally to the explicit user feedback and user task performance, we are interested in the system performance during a whole search session and how this relates to the user's feeling about system support. 
	\subsection{Apparatus}\label{subsec:apparatus}
	\paragraph{Prototypes}
	For the user study, we used the hotel search prototype introduced above -- in the following called weighted prototype. For the comparison with a conventional search interface using facets we developed a second prototype as a representative of a state-of-the-art hotel search interface (see Figure \ref{fig:facet_prototype}) -- in the following called facet prototype. This prototype provides filter functionalities in the form of facets, which are separated into different categories (see Figure \ref{fig:facet_prototype}a). For each facet term, the number of remaining results in the result list after choosing this facet is shown after the term in brackets. Selected facets are represented on the right-top side of the interface where the user can also delete selected facets by clicking on the recycle bin icon (see Figure \ref{fig:facet_prototype}b). The results shown in the result list (see Figure \ref{fig:facet_prototype}c) match all selected facet terms. The representation of the result list is the same as in the weighted prototype. However, an additional sorting function is available to sort the results by relevance, price, hotel stars and customer ratings (see Figure \ref{fig:facet_prototype}d). Both prototypes include a logging component to record user actions as well as the status of each presented result list. 
	
	\paragraph{Scenario}\label{paragraph:scenario}
	In order to compare the two prototypes and to analyze whole search sessions under the same conditions, we created a use case all participants had to perform with each prototype. The scenario includes "must-haves", "must-not haves" and "optional" preferences for a hotel in Berlin. The participants were asked to book a hotel for Paul who wanted to attend a conference in Berlin. His intended price range is between \euro{60} and \euro{120}. Participants were provided with Paul's preferences:
	"Paul would prefer to stay in district 'Mitte' (neighborhood) or in another district with good access to public transport (transport). By no means he wants to stay in the district 'Tiergarten' (neighborhood). In any case, he would like to have breakfast (meal). Furthermore, he would appreciate if the hotel has a restaurant (meal) or a bar (entertainment). Additionally, he would welcome, if the hotel has a fitness center (sport) and that he can pay on invoice (payment type). He could do without these two last items, whereas paying on invoice would be more important for him than the fitness center." The categories of the conditions were explicitly mentioned in the scenario because we are not interested in comparing how fast the participants could find the desired category compared to how fast they could type in the search terms. 
	There is no hotel in the data set that matches all criteria. Otherwise participants would find this hotel with both prototypes very quickly -- it would be the best hit after selecting all facets. In our scenario, Paul's preferences have to be weighted against each other and compromises have to be made finding a hotel that matches Paul's preferences properly. 
	
	\paragraph{Data sets and hotel relevance scores}\label{paragraph:dataset}
	We created a data set of 150 hotels in Berlin. The initial set of accommodations is based on information gained through the public Yelp-API. We received a list of hotels in Berlin with addresses, user comments, user ratings and categories. Each hotel was enriched by further features assigned to 18 different categories (such as price, neighborhood, type of room, type of catering, customer rating, etc.). This information was taken either from the hotel's website or information provided by Booking.com. To be able to have two comparable sets of hotels that can be used in the evaluation and to avoid a learning effect, we copied the hotel data set and changed only the names of the hotels. All information in the data set were provided in German.
	
	For evaluating how well a selected hotel fits Paul's preferences, we generated a graded relevance score ($grs$) for all hotels in our dataset. The algorithm is illustrated in Figure \ref{fig:algo}. First, we checked if the hotel match the "must have" / "must-not have" criteria. When these mandatory criteria are fulfilled the hotel gets a first $grs$ of one. In our case, that means, the hotel has a price between \euro{60} and \euro{120}, breakfast is included and the hotel is in the district "Mitte" and if it is not in "Mitte" it has access to public transport, and it is not in the district "Tiergarten". All other hotels that do not meet these criteria are considered to be not relevant and are not considered further. They get a $grs$ of zero. Then, we checked the additional optional criteria and awarded additional relevance scores depending on Paul's preferences. That means, a hotel with a fitness center gets two additional scores, as this was Paul's least preferred criterion. Hotels that provide the opportunity to pay by invoice gets three additional scores, as this was more preferred by Paul than the fitness center and hotels with a restaurant or a bar get four additional scores, as Paul preferred these features most. A hotel that meets all criteria can gain a $grs$ of 10 ($=1$ (mandatory criteria) $+2$ (fitness studio) $+3$ (paying on invoice) $+4$ (restaurant or bar)). In our data set only 15 hotels meet Paul's must-have preferences. No hotel meets all requirements, but five hotels have a graded relevance score of eight.
	
	%
	%
	\begin{figure}[]
	\begin{center}
	\begin{center}
	\includegraphics[trim=0 20 0 0,clip,width=0.9\linewidth]{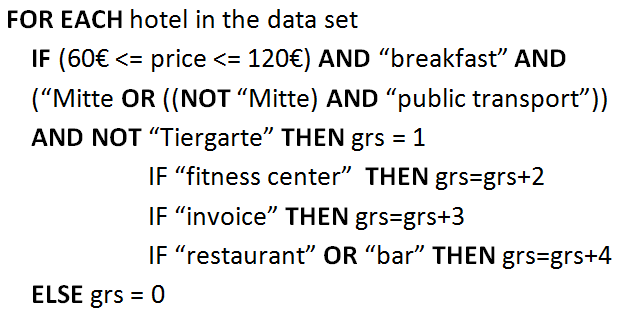}
	\end{center}
	\caption{Algorithm for calculating Hotel's graded relevance score (grs) based on Paul's preferences provided in the scenario.}
	\label{fig:algo}
	\end{center}
	\end{figure}
	
	\paragraph{Setup}
	For the user study, we used a laptop with internet access. Through the Firefox browser 46.0.1 both prototypes were accessed on our server. To make sure that all participants see exactly the same part of the user interface a 21\(^{\prime\prime}\) monitor with the same resolution was used in all sessions. Furthermore, an external keyboard and a mouse were provided to the participants as input modalities. The activities on the screen were recorded with the screen capture software Camtasia for further analysis. The used questionnaires were on paper.
	
	\subsection{Methodology}
	Data collection took place in a laboratory setting at a university and our institute in single sessions. We used a within-subject design approach with the two prototypes and the two data sets being the independent variables. The four resulting experimental conditions were randomly but equally assigned to the participants. The study took about 45 minutes per participant. As dependent variables, we collected time-on-task, clicks-on-task, graded relevance scores of selected hotels, subjective ratings, free-form text comments. Furthermore, we calculated the normalized discounted cumulative gain (NDCG) \cite{jarvelin2000ir} of each result list.
	
	\subsection{Procedure}
	The study was performed in single sessions and followed a detailed trail protocol with a counter-balanced order of the four experimental conditions. First, the experimenter explained the purpose of the study and that all activities on the screen are recorded. The participant agreed to the procedure by signing a consent form. Before the actual experiment started, the participant filled out a questionnaire in which we asked a few questions regarding demographic information and the experience with hotel search systems. Afterwards, the experimenter showed and explained the first prototype and asked the participant to familiarize herself with the interface for about four minutes. Then, the participant was given the assignment with the scenario description. The task ends with selecting a hotel for Paul. There was no time limit for the task. In a questionnaire, the participant was asked why she selected the hotel, how satisfied she is with her choice, how well the system supported her while searching for a hotel and if the ranking of the result list was comprehensible. The same procedure was followed with the second prototype. At the end of the study, the participant filled out a final questionnaire. In this, we asked for advantages, disadvantages of each prototype as well as for suggestions for improvements with open questions. Each participant received \euro{10} as a compensation for expenses. 
	
	\subsection{Participants}
	24 native speaking German participants took part in the user study. Half of them were university students from different fields of study. They were recruited through an announcement on notice boards or mailing lists. The other half were recruited by word-of-mouth recommendation and they all had either a university degree or a completed apprenticeship. 12 participants were female. Participants' age ranged from 18 to 51 years (M=28.04, SD=8.3). Three of them have never used a hotel search system. 16 of the participants are familiar with more than one hotel search system like Booking.com or Trivago.com and 17 used such a system at least once in the last six months. 81\% of these participants are satisfied or very satisfied with the functionalities such systems provide and 76\% are also satisfied or very satisfied with the search results of such systems.

	\section{Evaluation Results}
	In the following section, we describe the results of the user study by first focusing on the selected hotels, then on the provided user feedback and finally on system performance. Given the fact that all participants had to perform the task following a given scenario we are able to analyze changes in recall and precision over the whole search sessions and combine them with the provided user feedback on system support. We want to know if there are relations between user feedback and system performance measures. If not stated others, we use a Wilcoxon signed-rank test to compute differences between two paired groups with $\alpha \leq0.05$.
	
	\subsection{Analysis of selected hotels}
	
	\begin{table}[]
	\begin{center}
	\begin{center}
	\includegraphics[trim=50 680 130 55,clip, width=1.0\linewidth]{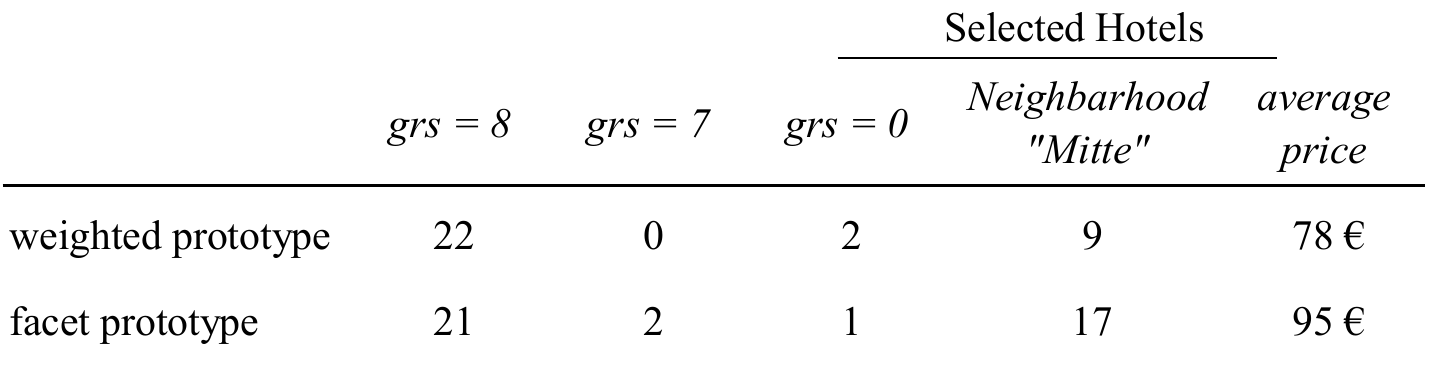}
	\end{center}
	\caption{Number of hotels selected by the participants for each prototype divided according to the graded relevance score, neighborhood "Mitte" and average price.}
	\label{table:hotelpoints}
	\end{center}
	\end{table}
	In almost all sessions, hotels with the highest possible graded relevance score were selected with both prototypes (see Table \ref{table:hotelpoints}). Having a closer look at the selected hotels, it is striking that with the facet prototype 17 participants chose a hotel in the neighborhood "Mitte", while with the weighted prototype only nine participants did the same. This might be an indication of a different level of elaborateness during the task performance. While the neighborhood was an equally weighted condition by Paul we found evidence in the log files that ten of the participants using the facet prototype did not consider that alternative at all. Furthermore, the log files showed that in total 16 participants had not examined all of Paul's preferences in the facet prototype. Beside of the preference "outside Mitte with access to public transport", the preference "fitness center" was often remained unconsidered. In the following, we will also have a closer look at those two user groups - those who examined all of Paul's preferences in the facet prototype (full examiners, n=8) and those who did not (incomplete examiners, n=16). 
	
	One interesting finding is the significant difference in price of the selected hotels (see Table \ref{table:hotelpoints}). With the weighted prototype, the accommodation price is significantly lower than with the facet prototype (facet M=99.63, weighted M=80.25, p=0.011). Whereby, in the group of full examiners there is no significant difference in the hotel price (facet M=85.82, weighted M=74). Nine participants also stated price as an important factor while searching for a hotel. 
	
	In the weighted prototype condition, two participants failed in solving the task correctly based on the provided information about the hotel. They chose a hotel outside the neighborhood Mitte without access to public transport.  However, one of them answered to our question why she chose this specific hotel with "There is good access to public transport." which indicates real knowledge about the neighborhood. With the facet prototype, also one participant chose a hotel outside Mitte with no access to public transport. We do not know if these two participants have the same reason (knowledge about the neighborhood) without stating it in the questionnaire. 18 participants provided further comments to the reasons why they chose a hotel. 15 stated that they included the user rating in their relevance decision process. For nine participants the price played an important role and comments from four participants allow the conclusion that they had further knowledge about Berlin, especially about the neighborhoods and the distances. 
	
	\subsection{Quantitative User Feedback}
	We asked the participants on five-point likert scales how satisfied they are with the selected hotel ("very satisfied"=1 to "not at all satisfied"=5), how well they felt supported in their search by the system ("very well"=1 to "not well at all"=5) and if the result sorting was comprehensible ("very comprehensible"=1 to "not at all comprehensible"=5). The results show that there is no significant difference regarding support and comprehension of both presented search systems. However, participants are significantly more satisfied with the selected hotel when they use the weighted prototype (M=1.67) than the facet prototype (M=2.13, Z=-2.082, p\textless.05). A closer look at our to user groups showed that this is true for the incomplete examiner (weighted M=1.63, facet M=2.19, p=0.039) but there was no significant difference in the group of full examiner. They were similar satisfied with the hotel selected in the weighted prototype (M=1.75) and in the facet prototype (M=1.88).
	
	For analyzing time-on-task and clicks-on-task needed to perform the task, we consulted the recorded screencast videos as well as the log file data. In both cases, we could find significant differences between the two prototypes. Participants are significantly faster and significantly fewer clicks are needed in the facet prototype than in the weighted prototype. In the facet prototype it took 5.3 minutes ($\sigma{=2.9}$) and 16.5 clicks ($\sigma{=10}$) on average to select a hotel compared to the weighted prototype with 7.36 minutes ($\sigma{=2.96}$) and 25.83 clicks ($\sigma{=9.76}$) on average (time-on-task: Z=-2.714, p\textless.05, clicks: Z=-3.244, p\textless.05). These results seems not surprisingly given the fact that the number of incomplete examiner is with 2/3 relatively high. In the group of full examiner, we could not find a significant difference in time-on-task (weighted prototype M=6.6 minutes, facet prototype M=7.52 minutes)  and clicks (weighted prototype M=18.13, facet prototype M=27.13). While in the group of incomplete examiner, the differences in time-on-task (weighted prototype M=7.64 minutes, facet prototype M=4.16 minutes) and clicks (weighted prototype M=24.69, facet prototype M=11.06) are significant (time: p\textless.0001; clicks: p=.000).
	
	\subsection{Qualitative User Feedback}
	In the final questionnaire, we collected qualitative user feedback in open questions on advantages, disadvantages and suggestions for both prototypes. In the following, we only provide feedback comments that were stated by more than one participant.
	
	\textit{Weighted prototype's advantages} 22 participants answered the question about advantages for the weighted prototype. Altogether we collected 34 different statements. The benefit most often stated, by eleven participants, was the opportunity to weight optional criteria. Five persons liked the "must-not have"-opportunity and also five found the interface well structured. Three participants appreciated the free-text search function and two other liked that they got more potential results that they can compare.
	
	\textit{Weighted prototype's disadvantages} 18 participants provided 28 disadvantage statements for the weighted prototype. Six persons disliked that there was no category list. Five missed an explicit sort-by function for price or ratings. Four were annoyed by typing in the search criteria. 
	
	\textit{Suggestions for the weighted prototype}
	We collected 18 suggestions for the weighted prototype, provided by 16 participants. Six persons would like to have an explicit sort function by price or ratings. Showing all available criteria was suggested by four participants. 
	
	\textit{Facet prototype's advantages}
	24 participants provided feedback to the question about advantages of the facet prototype, wherewith we collected altogether 30 statements. 15 persons liked to select the facet terms from lists shown on the left side of the interface, as they know it from online shops and travel portals. Four participants appreciated the opportunity to sort the results by a predefined sorting criterion. Two persons mentioned positively that they do not have to type in a search term and another two liked the clear arrangement. 
	
	\textit{Facet prototype's disadvantages}
	We received 24 statements from 19 participants to the question about disadvantages of the facet prototype. Six persons missed that there was no possibility to weight the criteria. For four participants the interface was too overloaded, and three criticized the cutting of hotels that did not fulfill all criteria. Two participants stated that it was cumbersome to find out which criteria led to a no-result list and two others missed the opportunity to compare alternatives that did not match all criteria.

	\textit{Suggestions for facet prototype}
	18 participants provided suggestions for the facet prototype. Altogether 21 comments could be collected of which eight would like to have a weighting function. Three would add a search field and two the possibility to exclude results matching a "must-not have" criterion. 

	\subsection{System Performance Measure}
	When analyzing the facet and the weighted prototype regarding system performance, we have to keep in mind that their individual functionalities influence recall and precision differently. In a faceted search system, recall is massively influenced by the filtering functionality of the facets. The weighting of facets, in contrast, mainly influences the precision. Concerning individual result lists, the two systems are hardly comparable. Therefore, we will evaluate them on their own on the level of the complete search session and combine results with user feedback. 
	
	\paragraph{Relevant hotels seen}
	In our context, recall would be defined as the ratio of the number of relevant hotels that the system retrieved for a user query and the total number of relevant hotels in the data set. However, in our study, the data set contained only 150 hotels. Stating queries with all relevant hotels within the result set is not a complex task. Instead of looking at the complete result list that a system retrieved, we will only consider the visible part of it, namely, all the results that have been displayed on the result pages the users browsed through. Also, we will not only focus on single queries, but we will incorporate all queries within a session. 
	
	Let us consider a participant's session in which multiple searches are conducted to find a suitable hotel. These queries can be triggered by entering search terms, filtering or reordering the result list, changing the weight of a criterion, and including or excluding criteria. Each query action will lead to a new result list of which the first 15 hotels are displayed on the first result page. These hotels form the visible part of the result list. If the participant proceeds to the next page of a result list, the number of visible results is increased. If we collect all visible hotels from each participant's search session, we can assess \textit{how many of the relevant hotels have been visible to the participants} using a specific system.
	
	\begin{figure}[]
	\begin{center}
	\begin{center}
	\includegraphics[width=1.0\linewidth]{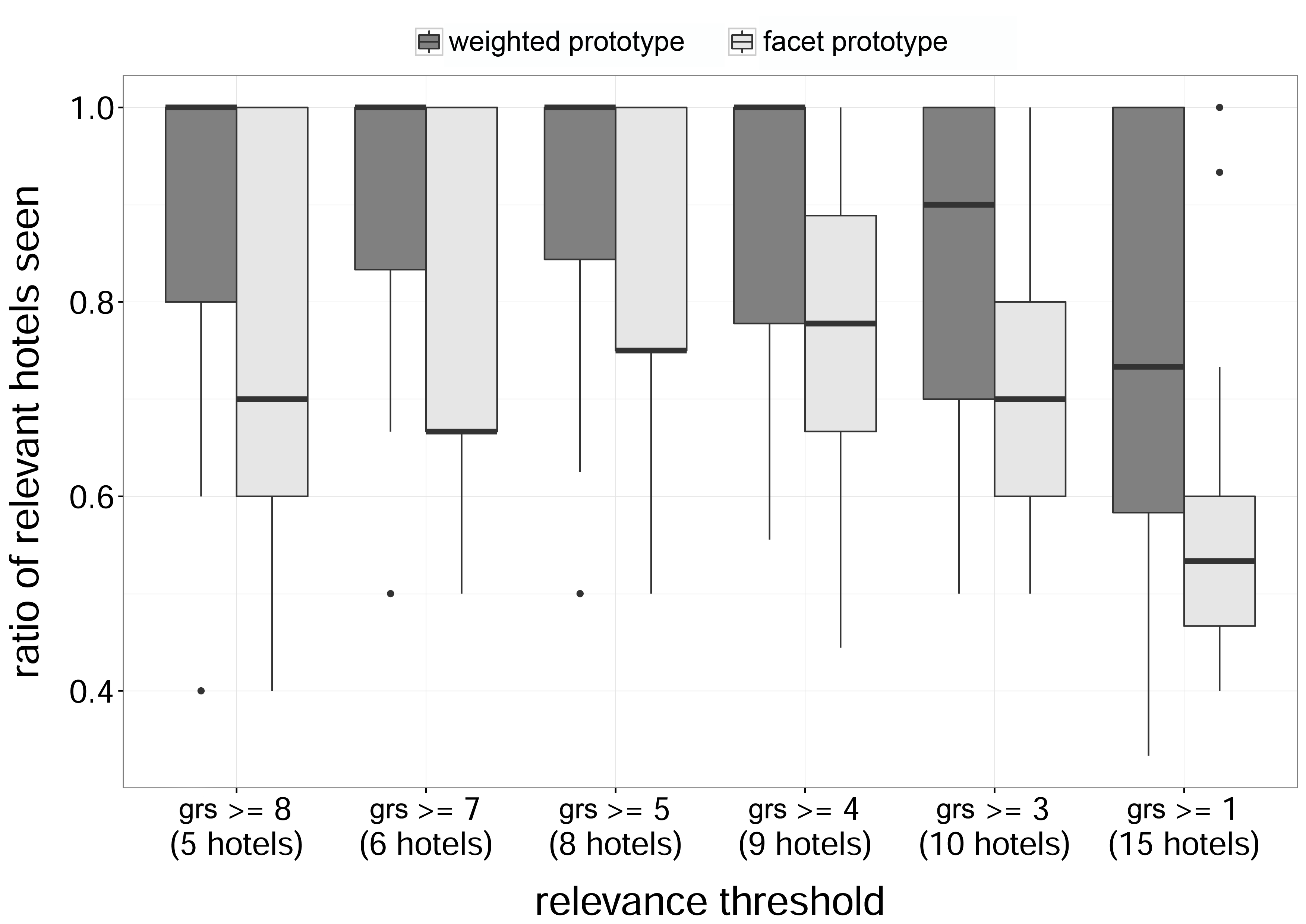}
	\end{center}
	\caption{Boxplots of the ratio between relevant hotels that were visible during all participants' sessions and the total number of relevant hotels, grouped by systems.}
	\label{fig:seen_hotels}
	\end{center}
	\end{figure}
	
	Figure \ref{fig:seen_hotels} shows boxplots of the ratio between relevant hotels that were visible during all participants' sessions and the total number of relevant hotels, grouped by systems. We calculated the ratio with a decreasing relevance threshold and grouped the results. The first two boxplots on the left show the ratio of hotels with a graded relevance score ($grs$) higher or equal 8 for the facet and the weighted prototype. The third and fourth boxplots show the ratio of visible hotels with a $grs$ higher or equal 7 and so forth. It can be observed that when using the weighted system until a relevance threshold of 4 a high mean of 1.0 and a first quartile of around 0.8 was achieved whereas the mean and first quartile using the facet prototype was lower. Overall, a non-parametric Mann-Withney test ($\alpha \leq0.05$) found the number of relevant hotels seen with the weighted system to be significantly higher. The results of the test are as follows: grs$\geq$8 $(u=2.668, p=.008)$, grs$\geq$7 $(u=2.645, p=.008)$, grs$\geq$5 $(u=2.544, p=.011)$, grs$\geq$4 $(u=2.900, p=.004)$, grs$\geq$3 $(u=3.099, p=.002)$, and grs$\geq$1 $(u=3.478, p=.001)$. Having a closer look at the two groups, it is not surprisingly that the group of incomplete examiner is again responsible for that result. They have seen significantly more hotels in the weighted prototype than in the facet prototype (for example grs$\geq$8 facet prototype M=3.50 vs. weighted prototype M=4.50, p=.002; and for grs$\geq$1 facet prototype M=7.94 vs. weighted prototype M=12.13, p\textless.0001). In the group of full examiner there are no significant differences. 

	\paragraph{Result list precision}
	For analyzing the quality of the result lists, we used the normalized discounted cumulative gain (NDCG) \cite{jarvelin2000ir}. Plotting the NDCG values of all searches conducted during a search session against time gives an overview of the development of the participant's session from the perspective of precision. As both systems' modes of operation have different impacts on the NDCG, the NDCG plots of the two systems should not be compared. Therefore, we will not compare the two systems concerning precision values, but we will investigate the relationship between precision and the perceived satisfaction with the systems' which was asked by the question "How well did you feel supported by the system?" ("very well"=1 to "not well at all"=5). 
	
	Figure \ref{fig:precision_weighted} shows the NDCG plots for the weighted prototype and Figure \ref{fig:precision_facet} for the facet prototype. Each point represents the NDCG value of a search at a specific point of time during the session. In both figures, the plots are grouped along the participants' answers to the question how well they felt supported by the system during their search. As the answers "not well" and "not well at all" were only given by one or none participant we only show the plots for "very well supported", "well supported" and "neither well nor not well supported".
	For example, in the upper plot of Figure \ref{fig:precision_weighted} ("very well"), each point represents the NDCG value of a search result list of a search session, where the participant felt very well (n=10) supported by the system. In addition to the data points, we generated LOESS smoothing curves\footnote{LOESS was first introduced by \cite{cleveland1979robust}. In this paper we use the implementation which is part of the R-Project: \url{https://stat.ethz.ch/R-manual/R-devel/library/stats/html/loess.html}} which helped to identify common characteristics. The dashed blue line is a LOESS curve created for the complete data set regardless of the participant's answer to the question of system support satisfaction. The solid red lines are a LOESS curve generated for the specific group.\footnote{All LOESS curves are generated with a degree of 2 and a span of 0.75.} In addition to the LOESS curves, for each plot, there exists a vertical dot dashed green line, which indicates the point of time, where the median of the group's participants has finished their task.
	
	\begin{figure}[]
	\begin{center}
	\begin{center}
	\includegraphics[trim=0 30 0 0,clip,width=.75\linewidth]{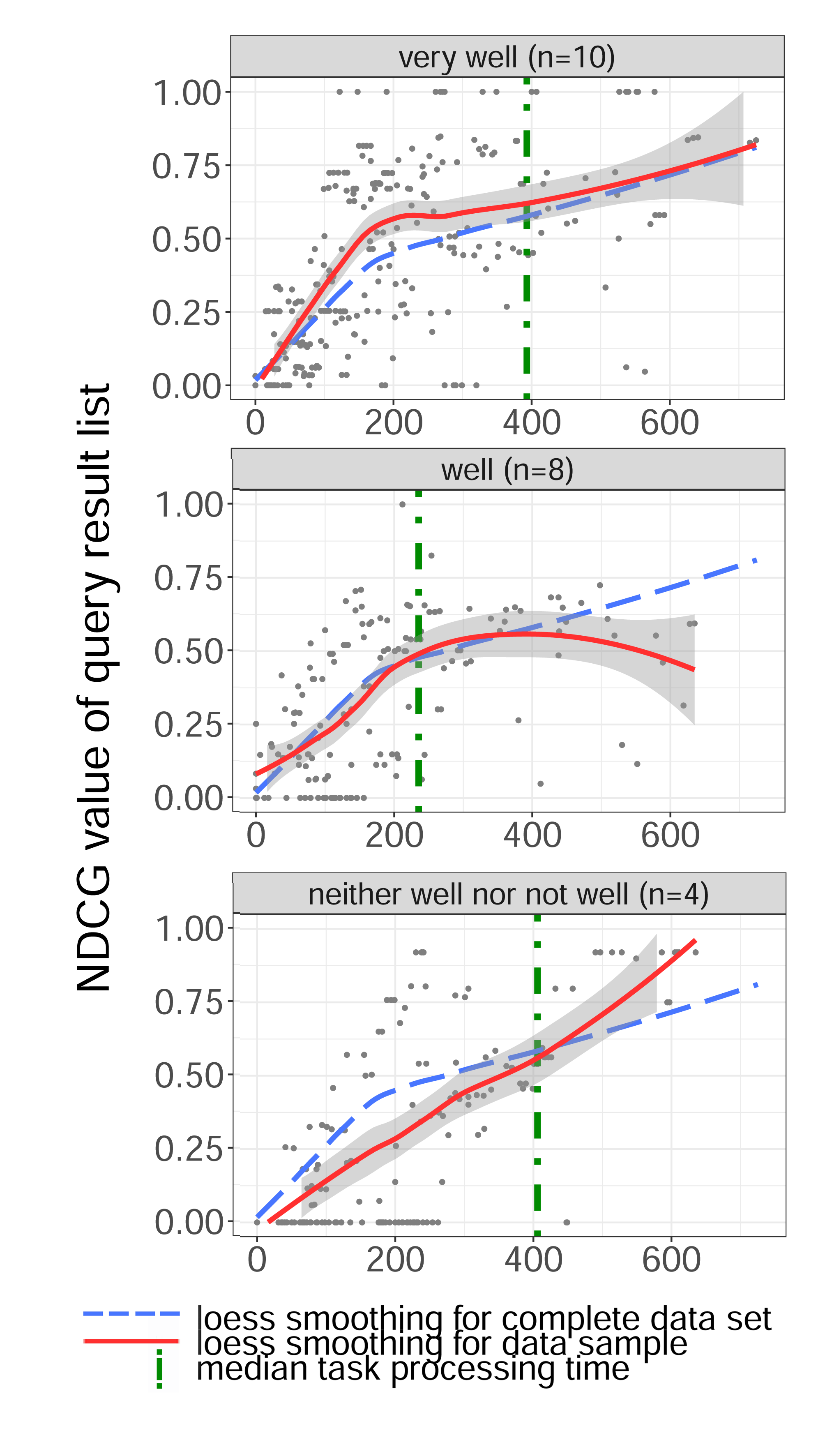}
	\end{center}
	\caption{NDCG plots for the weighted prototype, grouped along the answer to the question 'How well did you feel supported by the system?'.}
	\label{fig:precision_weighted}
	\end{center}
	\end{figure}
	
	\begin{figure}[]
	\begin{center}
	\begin{center}
	\includegraphics[trim=0 30 0 0,clip,width=.75\linewidth]{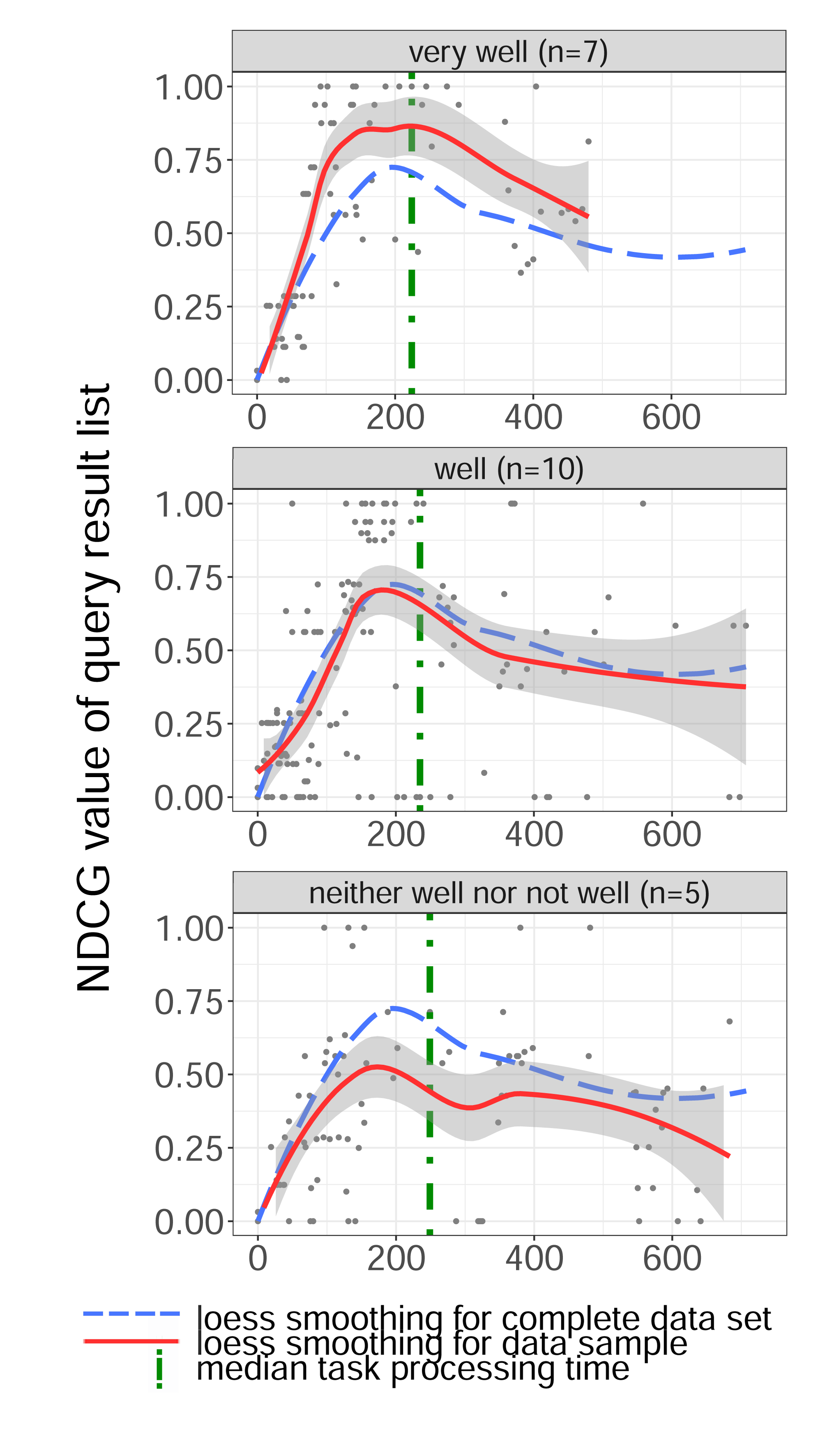}
	\end{center}
	\caption{NDCG plots for the facet prototype, grouped along the answer to the question 'how well did you feel supported by the system?'.}
	\label{fig:precision_facet}
	\end{center}
	\vspace{-3mm}
	\end{figure}
	
	The LOESS curves generated for the whole data set in Figure \ref{fig:precision_weighted} and Figure \ref{fig:precision_facet} show one common characteristic. The search process seems to be divided into two phases. First a take-off, where the precision of search requests increases until roughly 180 seconds into the session, where a break of slope indicates the beginning of the second phase. During the take-off phase, the steepness of the curve indicates a fast improvement of search precision. For the weighted prototype (Figure \ref{fig:precision_weighted}), this first phase ends at an NDCG value of around 0.4. After this, the NDCG still increases, but with a lower gradient. For the facet prototype (Figure \ref{fig:precision_facet}), the break of slope builds the maximum of an NDCG value of around 0.7. After the maximum, the curve slowly declines until a  minimum of around 0.4 after roughly 600 seconds. Notably, the break of slope lies before the median task processing time. This means that at least half of the participants (in most cases even 75\%) were still working on the task at the time point of the break of slope. This supports the assumption that the search sessions are indeed divided into two phases, as it means that the break of slope does not coincide with the completion of the task.
	
	Comparing the LOESS curve of the complete data sets and the curves of each individual subgroup one can observe a relationship between the satisfaction with the system's support and the precision of the search requests of the participants. In Figure \ref{fig:precision_weighted}, the LOESS curve of the group of participants that felt very well supported by the system primarily lies above the curve of the complete data set. Even if it is also divided into two phases, the take-off phase is steeper, and the break of slope lies on a higher level. Participants in this group were faster and better in formulating more precise search requests than the users in the other groups and felt better supported by the system. Similar results can be observed in the facet prototype condition.
	
	When comparing the individual groups' LOESS curves with the overall curves, we can identify similar relationships between those two for both systems. The lower the perceived level of support is, the lower the curve. The curve for the group "very well" is the highest, the curve for "well" is close to the overall curve, and the curve for "neither well nor not well" lies mostly underneath the overall curve. Overall, we can conclude that there seems to be a connection between perceived system support and the development of the search precision.
	
	\section{Discussion}
	In this paper, we compare two different concepts for searching products: (1) a pure facet concept, well established in all kinds of (product) search engines and (2) a preference-based approach using weighted facets which allows users to express preferences of certain product, in our case hotel features. We took user feedback, system performance and a combination of both into consideration to evaluate both approaches.  
	
	Our evaluation results show that the participants are significantly more satisfied with the selected hotels found in the weighted prototype than in the facet prototype. One possible explanation for that is given by the recall analysis which showed that there were significantly more relevant hotels visible during the whole search session. Therefore, participants were able to compare more hotels, even those that only partly fulfill their requirements, and might get a better feeling for their buying decision. This is inline with McSherry's findings that a decision maker wants to be informed of all items that are likely to be of interest \cite{McSherry2003}. The high number of relevant hotels shown is explainable by the different interaction techniques to compare alternative hotels. In the weighted prototype, based on the task, users themselves push more relevant hotels higher in the list with the interactive sliders, and therefore these hotels are better visible. In the facet prototype, the effort to see more relevant hotels is higher as the user has to select and deselect each facet and facet combinations explicitly to see its influence on the results. It seems that the incomplete examiner were not willing to take the extra effort. Apparently they are willing to select a hotel more quickly that matches at least parts of the preferences without knowing other, probably better, alternatives. Therefore, it is not surprising that these participants were faster and needed fewer clicks in the facet prototype to select a hotel. The time the full examiner took to find a hotel in the weighted and in the facet prototype did not differ significantly.
	Half of them were even faster in the weighted prototype than in the facet prototype. Furthermore, no participant complained in the questionnaire that finding a hotel with the weighted prototype, in general, took too much time or effort. The fact that with the weighted prototype the hotel price of the selected hotel is significantly lower than with the facet prototype could provide an incentive for some users to spend more effort for the search process. 
	
	Besides our comparison of the two prototypes regarding user satisfaction and recall, we were able to find similar characteristics within the search process of our participants by analyzing the search precision. Overall, the search sessions are divided into two phases. During the first phase, the take-off phase, the requirements defined in the scenario are transferred into the system, which leads to an increase in precision. During the second phase, the precision decreases and increases alternatively leading to a change of the precision curves slope. The reason here might be, that there is no perfect hotel for the task given and the participants had to change their queries to find alternatives that are close to the criteria given.
	
	Regarding result list quality we observed a relationship between the perceived system support and the precision of the participants. When plotting the NDCG values grouped by the participants answer to the question how well they felt supported by the system, one could observe, that a certain groups' precision curve lies above the average if the user felt very well supported and below if she felt neither well nor not well supported. This indicates that there might be a relation between the perceived system support and the precision of the participants' searches. However, this method is new, and we have not yet understood enough to draw solid conclusions, but we believe that the analysis of the search precision can aid in the task of measuring user satisfaction. In our case, the two prototypes do not allow for a comparison of the result list precision, as they generate those list differently. When comparing two similar systems, this method could produce comparable results, which would allow to evaluating two different versions of a system.
	
	One important lesson learned from our study have to be mentioned. Knowledge about Berlin, the city we chose as an example in our study, might have influenced the results, as participants selected hotels knowing that the neighborhood has a good connection to public transport. In further user studies, we will use a fictitious city. Furthermore, there are some other limitations in our approach which should not go unmentioned. The number of hotels in our data set with 150 hotels is rather low. We could not find an existing hotel data set with a sufficient number of facets for each hotel. So, we created our own by manually enhancing the dataset with a lot of different facets. However, we cannot preclude that the dataset size might have an effect on our evaluation results. In future user studies, we will use a data set with a higher number of hotels. In order to perform a combined analysis of user feedback with system performance it was necessary that all users performed the same task in both conditions. Also in this case, we cannot preclude that with a different task the results might be different. This is also an aspect we have to address in future research to examine further the relation between system performance and user satisfaction measures. 
	
	\section{Conclusions}
	In this paper, we present an evaluation of a search interface using a preference-based ranking approach.  Users can select, exclude and weight (optional) search criteria by their preferences and thus influence the ranking of the result list. In a user study, we compared this search interface to an interface using standards facets. 24 participants had the task to find a hotel according to predefined preferences with both search interfaces. We evaluated the interfaces from a user and a system performance perspective and found out that: 	
	\begin{itemize}
	\itemsep0pt
	\item Users are significantly more satisfied with the selected hotel found with the weighted prototype.
	\item Users were given more relevant hotels in the result lists with the weighted prototype.
	\item There is no significant difference regarding time-on-task and clicks when users examine all preferences in both prototypes.
	\item Users, who do not consider all preferences in their search queries in the facet prototype were significantly faster and needed fewer clicks to select a hotel than with the weighted prototype. 
	\item Users chose a significantly cheaper hotel with the weighted prototype.
	\item Both user interfaces show characteristic differences in the evolvement of precision during a search session.
	\item Users that were able to generate result lists with a higher precision seem to felt better supported by a system.
	\end{itemize}
	The last two results based on observations on the analysis of the relation
	between the participants' answers to the question how well they felt supported by the system and the precisions (measured by NDCG) of the result lists. In future work, we want to research the potential of analyzing the evolvement of search precision over whole search sessions as an indication for user satisfaction in more detail. 
	
	\section{Acknowledgement}
	We would like to thank Marco Janc for implementing the prototype. Additionally, we thank him and Maria Lusky for supporting us with the execution of the user study. We are also very grateful to those who participated in our study.

\bibliographystyle{SIGCHI-Reference-Format}
\bibliography{fuzzyfacets}


\begin{thebibliography}{00}


\ifx \showCODEN    \undefined \def \showCODEN     #1{\unskip}     \fi
\ifx \showDOI      \undefined \def \showDOI       #1{{\tt DOI:}\penalty0{#1}\ }
  \fi
\ifx \showISBNx    \undefined \def \showISBNx     #1{\unskip}     \fi
\ifx \showISBNxiii \undefined \def \showISBNxiii  #1{\unskip}     \fi
\ifx \showISSN     \undefined \def \showISSN      #1{\unskip}     \fi
\ifx \showLCCN     \undefined \def \showLCCN      #1{\unskip}     \fi
\ifx \shownote     \undefined \def \shownote      #1{#1}          \fi
\ifx \showarticletitle \undefined \def \showarticletitle #1{#1}   \fi
\ifx \showURL      \undefined \def \showURL       #1{#1}          \fi

\bibitem{bostandjiev2012tasteweights}
{Svetlin Bostandjiev}, {John O'Donovan}, {and} {Tobias H{\"o}llerer}. 2012.
\newblock \showarticletitle{TasteWeights: a visual interactive hybrid
  recommender system}. In {\em Proceedings of the sixth ACM conference on
  Recommender systems}. ACM, 35--42.
\newblock


\bibitem{Bruke}
{R.~D. Burke}, {K.~J. Hammond}, {and} {B.~C. Yound}. 1997.
\newblock \showarticletitle{The FindMe approach to assisted browsing}.
\newblock {\em IEEE Expert\/} {12}, 4 (Jul 1997), 32--40.
\newblock
\showISSN{0885-9000}
\showDOI{%
\url{http://dx.doi.org/10.1109/64.608186}}


\bibitem{carpineto_survey_2012}
{Claudio Carpineto} {and} {Giovanni Romano}. 2012.
\newblock \showarticletitle{A {Survey} of {Automatic} {Query} {Expansion} in
  {Information} {Retrieval}}.
\newblock {\em ACM Comput. Surv.\/} {44}, 1 (Jan. 2012), 1:1--1:50.
\newblock
\showISSN{0360-0300}
\showDOI{%
\url{http://dx.doi.org/10.1145/2071389.2071390}}


\bibitem{Chen2012}
{Li Chen} {and} {Pearl Pu}. 2012.
\newblock \showarticletitle{Critiquing-based recommenders: survey and emerging
  trends}.
\newblock {\em User Modeling and User-Adapted Interaction\/} {22}, 1 (2012),
  125--150.
\newblock
\showISSN{1573-1391}
\showDOI{%
\url{http://dx.doi.org/10.1007/s11257-011-9108-6}}


\bibitem{cleveland1979robust}
{William~S Cleveland}. 1979.
\newblock \showarticletitle{Robust locally weighted regression and smoothing
  scatterplots}.
\newblock {\em Journal of the American statistical association\/} {74}, 368
  (1979), 829--836.
\newblock


\bibitem{han2016user}
{Shuguang Han}, {Danchen Zhang}, {Daqing He}, {and} {Qikai Cheng}. 2016.
\newblock \showarticletitle{User exploration of slider facets in interactive
  people search system}.
\newblock {\em IConference 2016 Proceedings\/} (2016).
\newblock


\bibitem{Harper:2015:PUC:2792838.2800179}
{F.~Maxwell Harper}, {Funing Xu}, {Harmanpreet Kaur}, {Kyle Condiff}, {Shuo
  Chang}, {and} {Loren Terveen}. 2015.
\newblock \showarticletitle{Putting Users in Control of Their Recommendations}.
  In {\em Proceedings of the 9th ACM Conference on Recommender Systems} {\em
  (RecSys '15)}. ACM, New York, NY, USA, 3--10.
\newblock
\showISBNx{978-1-4503-3692-5}
\showDOI{%
\url{http://dx.doi.org/10.1145/2792838.2800179}}


\bibitem{Hearst:2002:FFW:567498.567525}
{Marti Hearst}, {Ame Elliott}, {Jennifer English}, {Rashmi Sinha}, {Kirsten
  Swearingen}, {and} {Ka-Ping Yee}. 2002.
\newblock \showarticletitle{Finding the Flow in Web Site Search}.
\newblock {\em Commun. ACM\/} {45}, 9 (Sept. 2002), 42--49.
\newblock
\showISSN{0001-0782}
\showDOI{%
\url{http://dx.doi.org/10.1145/567498.567525}}


\bibitem{jarvelin2000ir}
{Kalervo J\"{a}rvelin} {and} {Jaana Kek\"{a}l\"{a}inen}. 2000.
\newblock \showarticletitle{IR Evaluation Methods for Retrieving Highly
  Relevant Documents}. In {\em Proceedings of the 23rd Annual International ACM
  SIGIR Conference on Research and Development in Information Retrieval} {\em
  (SIGIR '00)}. ACM, New York, NY, USA, 41--48.
\newblock
\showISBNx{1-58113-226-3}
\showDOI{%
\url{http://dx.doi.org/10.1145/345508.345545}}


\bibitem{internetretailer:venue}
{Matt Lindner}. 2016.
\newblock "Online sales will reach \$523 billion by 2020 in the U.S.".
\newblock Online.   (29 January 2016).
\newblock
\newblock
\shownote{Retrieved September 20, 2016 from
  \url{https://www.internetretrailer.com/2016/02/29/online-sales-will-reach-523-billion-2020-us}.}


\bibitem{loepp2015blended}
{Benedikt Loepp}, {Katja Herrmanny}, {and} {J\"{u}rgen Ziegler}. 2015.
\newblock \showarticletitle{Blended Recommending: Integrating Interactive
  Information Filtering and Algorithmic Recommender Techniques}. In {\em
  Proceedings of the 33rd Annual ACM Conference on Human Factors in Computing
  Systems} {\em (CHI '15)}. ACM, New York, NY, USA, 975--984.
\newblock
\showISBNx{978-1-4503-3145-6}
\showDOI{%
\url{http://dx.doi.org/10.1145/2702123.2702496}}


\bibitem{McSherry2003}
{David McSherry}. 2003.
\newblock {\em Similarity and Compromise}.
\newblock Springer Berlin Heidelberg, Berlin, Heidelberg, 291--305.
\newblock
\showISBNx{978-3-540-45006-1}
\showDOI{%
\url{http://dx.doi.org/10.1007/3-540-45006-8_24}}


\bibitem{Parra:2014:SYW:2557500.2557542}
{Denis Parra}, {Peter Brusilovsky}, {and} {Christoph Trattner}. 2014.
\newblock \showarticletitle{See What You Want to See: Visual User-driven
  Approach for Hybrid Recommendation}. In {\em Proceedings of the 19th
  International Conference on Intelligent User Interfaces} {\em (IUI '14)}.
  ACM, New York, NY, USA, 235--240.
\newblock
\showISBNx{978-1-4503-2184-6}
\showDOI{%
\url{http://dx.doi.org/10.1145/2557500.2557542}}


\bibitem{Frei93effectivenessof}
{Hans peter Frei} {and} {Yonggang Qiu}. 1993.
\newblock Effectiveness of Weighted Searching in an Operational IR Environment.
\newblock   (1993).
\newblock


\bibitem{Pu:2005:ITS:1064009.1064038}
{Pearl Pu} {and} {Li Chen}. 2005.
\newblock \showarticletitle{Integrating Tradeoff Support in Product Search
  Tools for e-Commerce Sites}. In {\em Proceedings of the 6th ACM Conference on
  Electronic Commerce} {\em (EC '05)}. ACM, New York, NY, USA, 269--278.
\newblock
\showISBNx{1-59593-049-3}
\showDOI{%
\url{http://dx.doi.org/10.1145/1064009.1064038}}


\bibitem{Pu:2004:EES:988772.988804}
{Pearl Huan~Z. Pu} {and} {Pratyush Kumar}. 2004.
\newblock \showarticletitle{Evaluating Example-based Search Tools}. In {\em
  Proceedings of the 5th ACM Conference on Electronic Commerce} {\em (EC '04)}.
  ACM, New York, NY, USA, 208--217.
\newblock
\showISBNx{1-58113-771-0}
\showDOI{%
\url{http://dx.doi.org/10.1145/988772.988804}}


\bibitem{Ricci2011}
{Francesco Ricci}, {Lior Rokach}, {and} {Bracha Shapira}. 2011.
\newblock {\em Introduction to Recommender Systems Handbook}.
\newblock Springer US, Boston, MA, 1--35.
\newblock
\showISBNx{978-0-387-85820-3}
\showDOI{%
\url{http://dx.doi.org/10.1007/978-0-387-85820-3_1}}


\bibitem{rocchio_relevance_1971}
{J.~J. Rocchio}. 1971.
\newblock \showarticletitle{Relevance feedback in information retrieval}.
\newblock In {\em The {Smart} retrieval system - experiments in automatic
  document processing}, {G.~Salton} (Ed.). Englewood Cliffs, NJ: Prentice-Hall,
  313--323.
\newblock


\bibitem{Spyratos2006}
{Nicolas Spyratos} {and} {Vassilis Christophides}. 2006.
\newblock \showarticletitle{Querying with Preferences in a Digital Library}. In
  {\em Proceedings of the 2005 International Conference on Federation over the
  Web}. Springer-Verlag, Berlin, Heidelberg, 130--142.
\newblock
\showISBNx{3-540-31018-5, 978-3-540-31018-1}
\showDOI{%
\url{http://dx.doi.org/10.1007/11605126_8}}


\bibitem{stolze2000soft}
{Markus Stolze}. 2000.
\newblock \showarticletitle{Soft navigation in electronic product catalogs}.
\newblock {\em International Journal on Digital Libraries\/} {3}, 1 (2000),
  60--66.
\newblock
\showISSN{1432-5012}
\showDOI{%
\url{http://dx.doi.org/10.1007/PL00021475}}


\bibitem{tunkelang:doi:10.2200/S00190ED1V01Y200904ICR005}
{Daniel Tunkelang}. 2009.
\newblock \showarticletitle{Faceted Search}.
\newblock {\em Synthesis Lectures on Information Concepts, Retrieval, and
  Services\/} {1}, 1 (2009), 1--80.
\newblock
\showDOI{%
\url{http://dx.doi.org/10.2200/S00190ED1V01Y200904ICR005}}


\bibitem{Voigt:2012:WFB:2305484.2305509}
{Martin Voigt}, {Artur Werstler}, {Jan Polowinski}, {and} {Klaus Meissner}.
  2012.
\newblock \showarticletitle{Weighted Faceted Browsing for Characteristics-based
  Visualization Selection Through End Users}. In {\em Proceedings of the 4th
  ACM SIGCHI Symposium on Engineering Interactive Computing Systems} {\em (EICS
  '12)}. ACM, New York, NY, USA, 151--156.
\newblock
\showISBNx{978-1-4503-1168-7}
\showDOI{%
\url{http://dx.doi.org/10.1145/2305484.2305509}}


\bibitem{Wei:2013:SFS:2481562.2481564}
{Bifan Wei}, {Jun Liu}, {Qinghua Zheng}, {Wei Zhang}, {Xiaoyu Fu}, {and} {Boqin
  Feng}. 2013.
\newblock \showarticletitle{A Survey of Faceted Search}.
\newblock {\em J. Web Eng.\/} {12}, 1-2 (Feb. 2013), 41--64.
\newblock
\showISSN{1540-9589}
\showURL{%
\url{http://dl.acm.org/citation.cfm?id=2481562.2481564}}


\bibitem{Yee:2003:FMI:642611.642681}
{Ka-Ping Yee}, {Kirsten Swearingen}, {Kevin Li}, {and} {Marti Hearst}. 2003.
\newblock \showarticletitle{Faceted Metadata for Image Search and Browsing}. In
  {\em Proceedings of the SIGCHI Conference on Human Factors in Computing
  Systems} {\em (CHI '03)}. ACM, New York, NY, USA, 401--408.
\newblock
\showISBNx{1-58113-630-7}


\end{thebibliography}

\end{document}